# Single-photon Interference over 150 km Transmission Using Silica-based Integrated-optic Interferometers for Quantum Cryptography

Tadamasa Kimura, (*)Yoshihiro Nambu[1], Takaaki Hatanaka[2],

Akihisa Tomita[3], Hideo Kosaka[4] and Kazuo Nakamura[1]

Department of Materials Science and Engineering, Tokyo Institute of Technology, 4259 Nagatsuda, Yokohama, Kanagawa 226-0026, Japan

[1]Fundamental and Environmental Research Laboratories, NEC, 34 Miyukigaoka, Tsukuba, Ibaraki 305-8501, Japan

[2]Fiber Optic Devices Division, NEC Corp., 747 Magi, Ohtsukimachi, Ohtsuki, Yamanashi 401-0016, Japan

[3]Quantum Computation and Information Project, ERATO, JST, 34 Miyukigaoka, Tsukuba, Ibaraki 305-8501, Japan

[4]Research Institute of Electrical Communication, Tohoku University, 2-1-1 Katahira, Aoba-ku, Sendai, Miyagi 980-8577, Japan

(*) Corresponding author: E-mail address: y-nambu@ah.jp.nec.com

We have demonstrated single-photon interference over 150 km using time-division interferometers for quantum cryptography, which were composed of two integrated-optic asymmetric Mach-Zehnder interferometers, and balanced gated-mode photon detectors. The observed fringe visibility was more than 80% after 150 km transmission.

Quantum key distribution (QKD) allows two remote parties (e.g., Alice and Bob) to share a secret key, with privacy guaranteed by quantum mechanics [1-2]. Since the first demonstration of QKD over 30 cm in free space [2], extensive efforts have been devoted to extending the transmission distance using an optical fiber link and fiber-optic devices [3-10]. Time-division interferometers for optical pulses composed of one or two asymmetric Mach-Zehnder interferometers (AMZs) have been used to code random keys. A train of double pulses of a single photon carries bit data, where information is encoded in the relative phase between two pulses. Townsend et al. first demonstrated 10 km transmission of single interfering photons [3] using two AMZs. Later, the distance was extended to 48 km [5]. However, these systems were unstable because both the path length difference and the polarization characteristics of the two AMZs were unstable during operation. To overcome this problem, an autocompensating interferometer using a Faraday mirror was invented [7-9]. Recently, Kosaka et al. demonstrated single-photon interference over 100 km at a telecom wavelength [10] using a low-noise photon detector [11].

Although the autocompensating system works well for QKD systems using a faint pulse up to 100 km, extending the transmission distance will be difficult even if a lower noise photon detector is developed. This is because the backscattering noise in the fiber dominates the detector noise, which is intrinsic to the bidirectional autocompensating system [9]. In fact, high backscattering noise has been shown to be an obstacle to transmission over 100 km [10].

Although the use of storage line and burst photon trains would reduce the backscattering, this would also reduce the effective transmission rate by one-third [9].

In this letter, we propose a unidirectional system using integrated-optic interferometers based on the planar lightwave circuit (PLC) technology as a solution to this conflict between stability and transmission distance [12]. We have demonstrated single-photon interference with sufficiently high visibility for constructing a practical QKD system up to 150 km. Our system is also compatible with QKD systems using true single photon or quantum-correlated photon pairs.

AMZs with a 5 ns delay in one of the arms were fabricated on a silica-based PLC platform. This delay was sufficient to discriminate middle one of the three output pulses from the Bob's AMZ by a photon detector with a sub-nanosecond time window (see below). The optical loss was ~2 dB (excluding the 3 dB intrinsic loss at the coupler). Polarization-dependent loss was negligible (≤0.32 dB), but the birefringence of the waveguide could not be ignored. One of the couplers was made asymmetric to compensate for the difference in optical loss between the two arms. A Peltier cooler attached to the back of the substrate enabled control of the device temperature with up to 0.01°C precision. Polarization-maintaining fiber (PMF) pigtails aligned to the waveguide optic-axis were connected to the input and output of the AMZ.

Two AMZs were connected in series by the optical fiber to produce a QKD

interferometer system (Fig. 1). Optical pulses that were ≤200 ps long and linearly polarized along one of the two optic-axes were introduced into the PMF pigtail of Alice's AMZ from a gain-switched DFB laser at a wavelength of 1.55 μm. The input pulse was divided into two coherent output pulses polarized along the optic-axis of the output PMF, one passing through the short arm and the other through the long arm. The two optical pulses were attenuated so that their average number of photons μ was 0.2. The two faint pulses propagated along the optical fiber and underwent the same polarization transformation since the time scale of the polarization fluctuation in the fiber was much larger than the temporal separation between the two pulses. After traveling through Bob's AMZ, these pulses created three pulses in each of the two output ports. Among these three pulses, the first and last ones were independent of the relative phase between the two propagating pulses, whereas the middle one depended on the relative phase. Half of the photons received by Bob contributed to these interfering pulses.

To observe interference, both the relative phase setting of the two AMZs and the birefringence in the two arms of Bob's AMZ had to be controlled. This was achieved by controlling the device temperature. Since the AMZs were fabricated using the same mask, they had the same path length difference between the two arms, but their phase settings were not precisely determined. To set the phase, it is sufficient to control the path length difference within $\Delta L = \lambda/n$, where $n \sim 1.5$ is the refractive index of silica. The path length difference depends linearly on the device temperature with a proportionality constant ~5 μm/°C owing to the thermal expansion of the Si substrate. The birefringence in the two arms can be balanced by controlling the device temperature because the two arms have the same well-defined optic-axes on the substrate. Thus, it is sufficient to control the relative phase shift between two polarization modes to balance the birefringence. If the path length difference is a multiple of the beat length $\Delta L_B$ ($=\lambda/\Delta n$, where $\Delta n$ is the modal birefringence), the birefringence in the two arms is balanced and two pulses with the same polarization interfere with each other at the output coupler of Bob's AMZ no matter what the input pulse polarization is. Since $\Delta n/n$ was on the order of 0.01 for our device, the birefringence was much less sensitive to the device temperature than the relative phase. Therefore, we could easily manage both the phase setting and the birefringence balancing simultaneously.

Balanced, gated-mode InGaAs/InP avalanche photodiodes (APDs) were used to detect a single photon [10, 11]. The differential signal of the two APDs was recorded to eliminate common transient noise accompanying the gating operation. Two discriminators, which respectively responded to the positive and negative pulses, distinguished which APD was fired. Using this specially designed detector, we could achieve photon detection with a high S/N ratio. When the gate pulse width was 750 ps, the dark count probability was $2.1 \times 10^{-7}$ at a quantum efficiency of 10% per gate at $-108$°C. The interfering signal at the middle pulses was discriminated by adjusting the applied gate pulse timing. The system repetition rate was 1

MHz to avoid the APD after pulsing.

We measured the photon counting probability given by the key generation rate divided by the system repetition rate and plotted it as a function of transmission distance (Fig. 2). The measured data fit well with the upper limit determined by the loss of the fiber used (≈0.22 dB/km). A dispersion-compensating fiber was used to avoid photon count loss, which happens when the temporal broadening of the photon arrival time after long-distance transmission is larger than the time window of the photon detectors. However, we can avoid the need for such a special fiber using a sufficiently narrow spectral light source. In Fig. 2, the base lines show the dark count probabilities. Also, the interference fringe after 150 km transmission is shown in the inset. The observed visibilities were 82% and 84% for the two APDs. These figures were smaller than the theoretical upper limit determined by the ratio of the photon-count probability and the dark-count probability. This was due to a small short-term temperature drift just after the device temperature was changed to change the phase; the temperature drift caused a small phase drift and decreased visibility. This phase drift can be eliminated using an external phase modulator as shown below. The interference was stable over a long period without any active feedback control other than temperature control of the two independent AMZs, which is desirable for a QKD system. This shows that our system has the potential to achieve a much longer transmission distance than that attained in a previous experiment using the autocompensating system [10].

Although this letter reports only on single-photon interference over 150 km using temperature-dependent phase modulation of AMZs, a BB84 QKD system [1] should be feasible by inserting two fiber-optic phase modulators, one in each of the Alice's and Bob's apparatus, into the optical fiber link and selectively applying modulation to one of the propagating double pulses. However, the insertion of the extra devices might increase the loss of Bob's apparatus and the quantum bit error rate (QBER). To solve this problem, we recently proposed a BB84 QKD system that eliminates the need for the phase modulator in Bob's apparatus [13]. The observed visibility provides a direct measure of quantum bit error rates (QBERs) for this system. The QBERs estimated by QBER=$(1-V)/2$ [10] after 150 km transmission were 9% and 8%, respectively. The present result suggests that the PLC interferometer enables us to construct a practical QKD system up to 150 km using the system given in ref. 13.

In summary, we have realized backscattering-noise-free, 150-km-long single-photon interference through a unidirectional QKD system using integrated-optic interferometers. The observed interference was stable over a long period and fringe visibility was more than 80% after 150 km transmission.

This work was partly supported by the Telecommunications Advancement Organization of Japan.

**Figure Captions**

Fig. 1. Schematic diagram of the integrated-optic interferometer system. LD: laser diode, ATT: attenuator, APD: avalanche photodiode, DS: discriminator, CT: counter, H: 180° hybrid junction.

Fig. 2. Photon-counting probability as a function of transmission distance. Open triangles indicate the experimental results obtained from ref. 10. Inset: Fringe observed in photon count rate, obtained by changing the device temperature at 150 km.

Figure 1

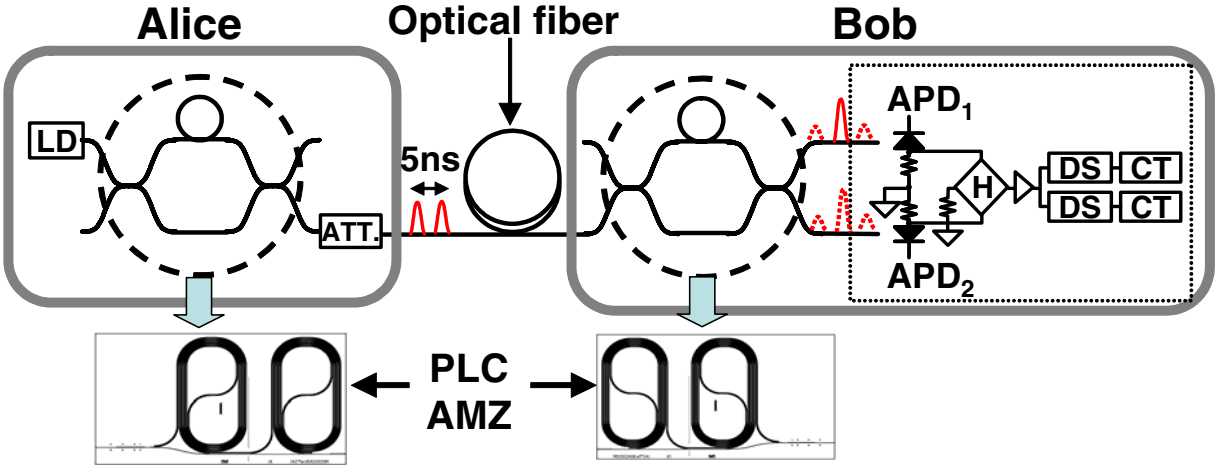

Figure 2

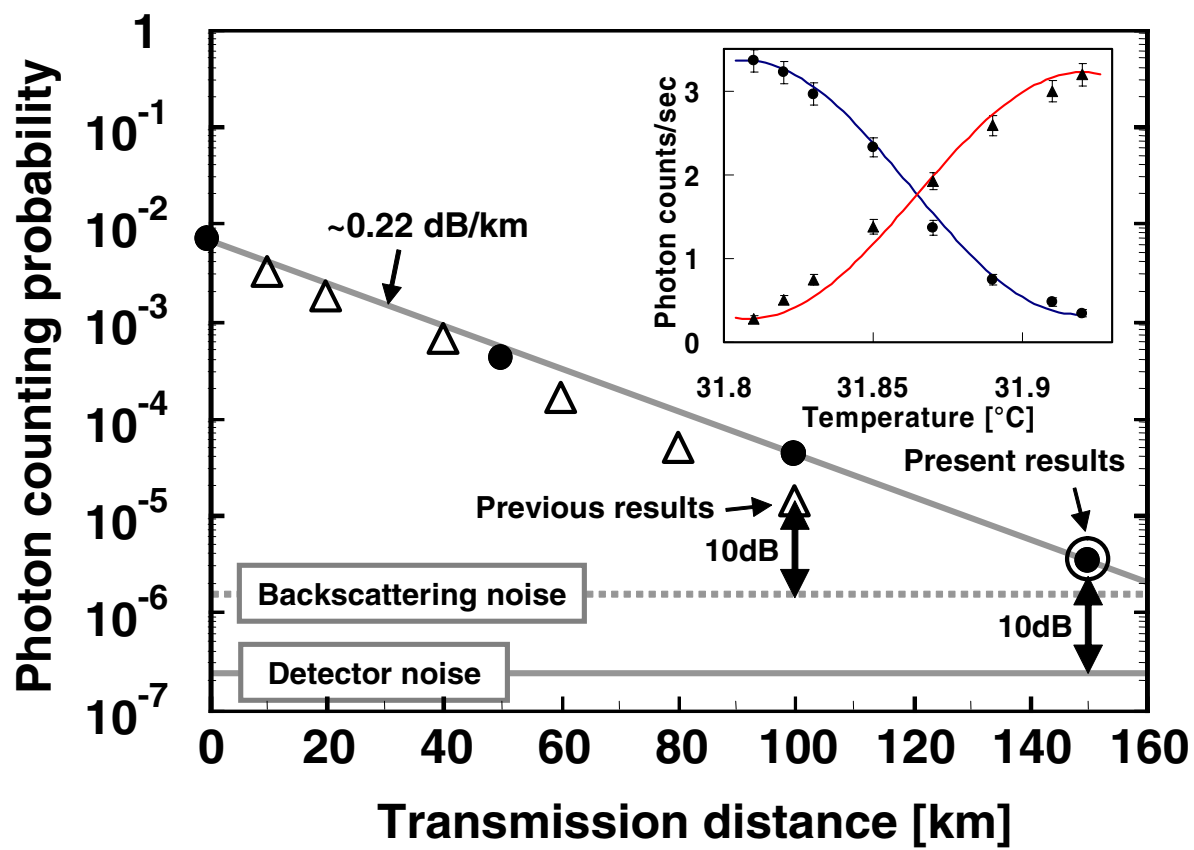